\documentclass[a4paper, 11pt]{article}
\usepackage{jheppub}
\usepackage{appendix} 
\newcommand{\tr}{\,\mathrm{tr\,}}

\title{On four dimensional $\mathbf{N=3}$ superconformal theories}
\preprint{WIS/11/15-NOV-DPPA}

\author{Ofer Aharony and Mikhail Evtikhiev
\\
\it{Department of Particle Physics and Astrophysics,\\
Weizmann Institute of Science, Rehovot 7610001, Israel}
}
\emailAdd{Ofer.Aharony@weizmann.ac.il}
\emailAdd{Mikhail.Evtikhiev@weizmann.ac.il}
\abstract{
In this note we study four dimensional theories with $N=3$ superconformal symmetry, that do not also have $N=4$ supersymmetry. No examples of such theories are known, but their existence is also not ruled out. We analyze several properties that such theories must have. We show that their conformal anomalies obey $a=c$. Using the $N=3$ superconformal algebra, we show that they do not have any exactly marginal deformations preserving $N=3$ supersymmetry, or global symmetries (except for their R-symmetries). Finally, we analyze the possible dimensions of chiral operators labeling their moduli space.
}

\begin{document}

\maketitle

\section{Introduction and summary of results}

Supersymmetric theories have been extensively studied in the past 40 years. This has partly been because their enhanced symmetries allow for various exact computations to be performed in these theories, even at strong coupling, so that they provide useful windows into strong coupling physics. In particular in four space-time dimensions, much progress has been made on understanding the properties of four dimensional $N=1$, $N=2$ and $N=4$ superconformal field theories (SCFTs).

In this note we study the properties of general $N=3$ superconformal theories, that do not also have $N=4$ supersymmetry -- we will call these ``pure $N=3$ theories''. There are no known examples of such theories, but to the best of our knowledge there is also no proof that they do not exist. The only free multiplet of $N=3$ supersymmetry is a vector multiplet identical to that of $N=4$ theories, so there are no free pure $N=3$ theories. There are also no weakly coupled $N=3$ SCFTs, since weakly coupled $N=2$ SCFTs are specified by their field content, and any such theory that has $N=3$ supersymmetry also has $N=4$ supersymmetry. So, all pure $N=3$ theories must be strongly coupled.

Various methods to construct strongly coupled $N=3$ theories have not yet yielded any examples of such theories. No brane construction in string theory \cite{GK} gives rise to such a theory. There are no known $AdS_5$ backgrounds of string or M theory, or of their low-energy supergravities, that have precisely $4d$ $N=3$ superconformal symmetry, though there is also not yet any proof that such backgrounds do not exist\footnote{We thank G. Papadopoulos for a discussion of these topics.}. One possible way to obtain $N=3$ theories would be to start from an $N=4$ theory and to perform a relevant deformation that preserves $N=3$ supersymmetry but not $N=4$, but we show in appendix \ref{relevant} that this is not possible. A large class of $N=2$ SCFTs arises (following \cite{Gaiotto}) by compactifying six dimensional $N=(2,0)$ superconformal theories on Riemann surfaces, but all examples of this type that have $N=3$ supersymmetry also have $N=4$. Finally, it may be possible to study $N=3$ superconformal theories by conformal bootstrap methods, imposing $N=3$ supersymmetry and a gap in dimensions between the conserved supercurrents and the next operator of spin $3/2$, but this has not yet been done (a conformal bootstrap analysis for $N=4$ theories was performed in \cite{Beem:2013qxa}, and for $N=2$ theories in \cite{Beem:2014zpa}).

Thus, it is interesting to either find examples of pure $N=3$ theories, or to prove that such theories cannot exist. In this note we take first steps towards this, by analyzing some of the properties that such $N=3$ theories must have. We begin in section \ref{anom} by proving that any $N=3$ theory must have a relation $a=c$ between its conformal anomalies, like $N=4$ SCFTs (but unlike general $N=2$ theories, whose ratio $a/c$ is bounded \cite{Hofman:2008ar} but not determined). Our method is to use the fact that any $N=3$ theory is also an $N=2$ theory, and to use the known facts about $N=2$ anomalies to learn about the $N=3$ anomalies. 

In section \ref{properties} we discuss three separate properties of pure $N=3$ theories. 
We show that pure $N=3$ theories cannot have exactly marginal deformations preserving $N=3$ supersymmetry, even though such deformations exist in all $N=4$ theories, and they are common in $N=1$ and $N=2$ theories (see \cite{Green:2010da} and references therein). Thus it is impossible to have a family of $N=3$ superconformal theories. This could have been expected from the fact that there are no weakly coupled $N=3$ theories, and one might expect any manifold of $N=3$ theories to have a weak coupling limit. We then show that pure $N=3$ theories, like $N=4$ theories, cannot have global symmetries that are not R-symmetries.

Finally, we analyze the chiral operators labeling the moduli space of $N=3$ theories (which is always a ``Coulomb branch'', since the only free multiplet is a vector multiplet). We find that these always include operators similar to the ones labeling moduli spaces of $N=4$ theories, with integer scaling dimensions. In pure $N=3$ theories their dimensions must obey $\Delta \geq 3$. However, we could not prove that additional operators are not also needed to parameterize the full moduli space.

It would be interesting to study additional constraints on $N=3$ theories, for instance by further constraining their Coulomb branches and attempting to classify at least $N=3$ theories which have Coulomb branches, as in for instance \cite{Argyres,Argyres:2005wx,Shapere,Argyres:2010py, Argyres:2015ffa}. One could also try to study them using the conformal bootstrap, or using the relation of general $4d$ $N=2$ theories to chiral algebras in two dimensions \cite{Beem:2013sza}.  It would also be interesting to generalize our analysis to different numbers of supercharges and dimensions.

{\it Note added:} After we submitted this paper, a paper \cite{Garcia} appeared, where several examples of $N=3$ theories were constructed. The results obtained there seem to be consistent with our statements.

\section{Conformal anomalies of $\mathbf{N=3}$ superconformal theories}\label{anom}

Let us begin by computing the conformal anomalies of $N=3$ superconformal theories. As usual in superconformal theories, the conformal anomalies $a$ and $c$ appearing in the trace of the energy-momentum tensor in curved space are related by supersymmetry to the chiral anomalies of the R-symmetry currents \cite{Anselmi}. In the case of $N=3$ SCFTs the R-symmetry group is $SU(3)_R\times U(1)_R$. We will denote the $U(1)_R$ charge by $R_3$; the supercharges are in the $\bf{3}$ representation of $SU(3)_R$ and have $R_3=1$. We would like to relate the R-symmetry anomalies to the conformal anomalies, and to see if this implies any restriction on the conformal anomalies. One way to do this would be to analyze in detail the general form of the 3-point function of the $N=3$ energy-momentum multiplet, which includes the energy-momentum tensor and the R-symmetry currents. However, since this is cumbersome, we will not do this here. Instead, we will see that the answer can be found simply by using the fact that $N=3$ theories are also $N=2$ 
theories, and using known facts about $N=2$ SCFTs.

The $N=3$ $SU(3)_R$-symmetry group generators can be expressed in the usual $SU(3)$ basis $T_a = \frac{\lambda_a}{2}$, where $\lambda_a$ are the Gell-Mann matrices, obeying $\tr(T_a T_b) = {1\over 2} \delta_{ab}$; two Cartan generators are $T_3$ and $T_8$ ($T_3 = \frac{1}{2}\; {\rm diag}(1,-1,0)$, $T_8 = \frac{1}{2\sqrt{3}}\; {\rm diag} (1,1,-2)$). $N=2$ superconformal theories have R-symmetry group  $SU(2)_R\times U(1)_{\tilde R}$; we will denote the $N=2$ $U(1)$ charge by $R_2$, and normalize it such that the $N=2$ supercharges have $R_2=1$.

When we view the $N=3$ theory as an $N=2$ theory, the $SU(2)_R\times U(1)_{\tilde R}$ symmetry of the latter is embedded into $SU(3)_R\times U(1)_R$; the other generator of $SU(3)_R\times U(1)_R$ that commutes with the $N=2$ supercharges is a global symmetry from the point of view of the $N=2$ theory. We will embed the $SU(2)_R$ in the top $2\times 2$ block of $SU(3)_R$, so that its Cartan generator is $I_3 = T_3$. The $R_2$
generator must then take the form $R_2 = \kappa R_3 + \mu T_8$. To get the correct $R_2$ charge for the $N=2$ supercharges, we must have 
$\kappa + \frac{\mu}{2\sqrt{3}} = 1$. To obtain another relation between the coefficients, we use the fact that (for a given choice of the two $N=2$ supercharges) the $N=2$ reduction of an $N=3$ theory is unique, so it is enough to determine $\kappa$ and $\mu$ in a specific example. In the $N=4$ theory, there is just an $SU(4)$ R-symmetry, that should include both $U(1)$ R-symmetries above; since both $R_3$ and $R_2$ must correspond to traceless generators, the R-charges of the $N=4$ supercharges must be $R_3 = (1,1,1,-3)$ and $R_2=(1,1,-1,-1)$, respectively. Comparing the charges of the third supercharge then gives $\kappa -2 \frac{\mu}{2\sqrt{3}} = -1$, leading to $\kappa=\frac{1}{3}, \mu =\frac{4}{\sqrt{3}}$. So, the $N=2$ R-symmetry is embedded into the $N=3$ R-symmetry as
\begin{equation} \label{chargerel}
I_3 = T_3,\qquad\qquad R_2 = \frac{1}{3} R_3 + \frac{4}{\sqrt{3}} T_8.
\end{equation}

The $N=3$ theory has three independent R-current cubic chiral anomalies: $SU(3)^3$, $SU(3)^2 \times U(1)$ and $U(1)^3$. There is also an anomaly for the $U(1)_R$ current with two energy-momentum tensors, that we will not require here. We normalize the anomalies using their value in a free theory as a trace over Weyl fermions  : 
\begin{equation} 
s_3 \equiv \tr(T_8^3) ,\qquad\qquad  s_2 \equiv \tr(R_3 T_3^2), \qquad\qquad r_3 \equiv \tr(R_3^3).
\end{equation} 
The structure constant relations for Gell-Mann matrices imply that 
\begin{equation} \label{strucrel}
\tr(T_3^3) = \tr(T_8^2 T_3) = \tr(R_3 T_3 T_8) = 0,\qquad \tr(T_8 T_3^2 )= -s_3,\qquad \tr(R_3 T_8^2) = s_2.
\end{equation}

In order to compute the $N=3$ anomalies and constrain $a/c$,  we use the supersymmetry reduction to $N=2$, which turns out to be enough. Two equations can be obtained by considering the form of the cubic anomalies in $N=2$ superconformal theories (see e.g.~\cite{Shapere}, following the $N=1$ results of \cite{Anselmi}), and using \eqref{chargerel}:
\begin{align}
&\tr(R_2^3) =  48(a-c) \qquad \rightarrow \qquad \frac{1}{27}r_3 + \frac{16}{3} s_2 + \frac{64}{3\sqrt{3}}s_3 = 48(a-c),\label{aconst1}\\
&\tr(R_2 I_3^2) = 4a-2c \qquad \rightarrow \qquad \frac{1}{3}s_2 - \frac{4}{\sqrt{3}}s_3 = 4a-2c.\label{aconst2}
\end{align}

To proceed further, we can use the extra $U(1)$ global symmetry that we get (from the $N=2$ point of view) from the $N=3$ R-symmetry. Since the $N=2$ supercharges cannot be charged under this global symmetry, its form must be (up to an unimportant overall factor) $F = R_3 - 2 \sqrt{3} T_8$.
According to equation (3.33) of \cite{Kuzenko}, the three-point function of two $N=2$ supercurrents and a flavor current superfield vanishes.
In particular the parity-odd term in the 3-point function of the currents vanishes. This term gets contributions
from both the $U(1)_{\tilde R}$ and $SU(2)_R$ currents in the $N=2$ supercurrent. Being careful about the normalizations, the two terms that contribute are
\begin{align}
	&U(1): \qquad \tr(R_2 R_2 F) \Rightarrow \frac{1}{4}\tr((R_3 -2\sqrt{3} T_8) (\frac{1}{3} R_3 + \frac{4}{\sqrt{3}} T_8)^2),  \label{aconst31}\\
	&SU(2): \qquad \tr(I_3 I_3 F) \Rightarrow 3 \tr ((T_3^2) (R_3 -2\sqrt{3} T_8)),  \label{aconst32}
\end{align} 
where the factor of $3$ on the second line comes from summing over the $3$ $SU(2)_R$ generators, and the factor of $\frac{1}{4}$ on the first line from normalizing $R_2$ in the same way as the $SU(2)_R$ generators.
Adding up \eqref{aconst31} and \eqref{aconst32} we get
\begin{equation}
	\frac{1}{9}r_3 + 12 s_2 + \frac{40}{\sqrt{3}}s_3 = 0. \label{aconst3}
\end{equation}

We now have $3$ equations for $r_3$, $s_2$ and $s_3$, which are enough to determine them, but not enough to give
constraints on $a$ and $c$. However, there is another restriction on the anomalies coming from (3.32) of \cite{Kuzenko}. This equation shows that the correlator of three global symmetry currents in $N=2$ SCFTs is antisymmetric, and thus it should vanish for three identical currents, so that
\begin{equation}
	\tr(F^3)  = r_3 -24\sqrt{3}s_3 + 36s_2  = 0. \label{atoc}
\end{equation} 
Altogether we now have $4$ equations
\begin{align}
	& \frac{1}{27}r_3 + \frac{16}{3} s_2 + \frac{64}{3\sqrt{3}}s_3 = 48(a-c),\\
	& \frac{1}{3}s_2 - \frac{4}{\sqrt{3}}s_3 = 4a-2c,\\
	& \frac{1}{9}r_3 + 12 s_2 + \frac{40}{\sqrt{3}}s_3 = 0,\\
	& r_3 + 36s_2-24\sqrt{3}s_3   = 0.
\end{align}
These equations only have a solution if $a=c$, and then we find
\begin{equation}
	r_3 = -96 c, \qquad
	s_2 = 2c, \qquad
	s_3 = -\frac{c}{\sqrt{3}}.
\end{equation}

Thus, as in $N=4$ theories, $N=3$ SCFTs necessarily have $a=c$. Note that since our analysis is valid for any $N=3$ theory, including as a special case the $N=4$ theories, it is obvious that any relation that we find between $a$ and $c$ must take the form $a=c$. Our non-trivial result is that indeed the constraints on the anomalies give such a relation, and thus lead to $a=c$. It would be interesting to see this also from a direct analysis of the 3-point functions of the $N=3$ energy-momentum supercurrent.

\section{Properties of $\mathbf{N=3}$ theories}\label{properties}

In this section we study some general properties of 
 $N=3$ superconformal theories that are not also $N=4$ (we call these ``pure $N=3$ theories'').
 In section \ref{emd} we show that pure $N=3$ theories cannot have any exactly marginal deformations, so that any such theories are isolated fixed points. In section \ref{cvr} we show that these theories, like $N=4$ theories, cannot have any global symmetries. Finally, in section \ref{cbos} we discuss the short multiplets whose expectation values can parametrize Coulomb branches of $N=3$ SCFTs.
Throughout this section we use several facts about $N=2$ superconformal multiplets, that can be found in \cite{Dolan}.

\subsection{$\mathbf{N=3}$ multiplets with exactly marginal deformations} \label{emd}

An exactly marginal deformation is a scalar operator of dimension 4 that sits at the top of a supermultiplet and is invariant under R-symmetry. All $N=4$ theories have a complex exactly marginal deformation sitting in their energy-momentum multiplet, so let us first analyze this case, viewing this theory as a special case of an $N=3$ superconformal theory. The bottom state of the $N=4$ energy-momentum multiplet is a dimension $2$ scalar belonging to the ${\bf 20'}$ representation of $SU(4)$; it breaks into ${\bf 6}_{-4}$, ${\bar {\bf 6}}_{4}$ and ${\bf 8}_0$ representations of $SU(3)_R\times U(1)_R$. Since all of these representations have different $U(1)_R$ charges for their bottom components, they must give birth to different $N=3$ multiplets (such multiplets are built on a bottom component that has a specific spin and R-symmetry representation). Thus, the $N=4$ energy-momentum multiplet splits exactly into three $N=3$ multiplets. 

The multiplet built on ${\bf 8}_0$ is the energy-momentum multiplet of the $N=3$ theory, which is always present. However, the complex exactly marginal deformation belongs to the multiplets built on ${\bf 6}_{-4}$ and ${\bar {\bf 6}}_{4}$; this can be seen from the fact that it is obtained from the bottom component by the action of either 4 $Q$'s or 4 $\bar{Q}$'s, and it should have $R_3=0$. Under the reduction of $N=4$ to $N=3$, three of the four supercharges sit in the $N=3$ energy-momentum multiplet, while the fourth necessarily belongs to the multiplets built on ${\bf 6}_{-4}$ and ${\bar {\bf 6}}_{4}$. Thus, the exactly marginal
deformation of $N=4$ sits in the same $N=3$ representation as an extra supercurrent, and cannot appear in pure $N=3$ theories.

Now let us analyze the general case. Exactly marginal deformations of any $N=2$ SCFT sit in chiral representations (denoted ${\cal E}_{2(0,0)}$) whose bottom component is a scalar of dimension $\Delta=2$, that is an $SU(2)_R$ singlet with $R_2 = \pm 4$. So it is clear that the bottom component of an $N=3$ multiplet containing an exactly marginal deformation must obey $\Delta_3 \leq 2$. On the other hand, the dimension must be a half-integer, and it cannot be equal to $\frac{3}{2}$ (since a spinor operator of this dimension is a free field) or $1$ (since a scalar operator of this dimension is a free field). 
So the bottom component of the multiplet should have $\Delta_3=2$, and contain a scalar singlet of $SU(2)_R$ with
$R_2 = \pm4$. For deformations preserving $N=3$ supersymmetry, the top component must have $R_3=0$, so it is clear that this bottom component must also have $R_3 = \pm4$. Using the unitarity constraints of appendix \ref{scaldim} on $N=3$ superconformal representations, and the decomposition of $SU(3)$ to $SU(2)\times U(1)$, the only
possible $SU(3)_R$ representations with $R_3 = \pm 4$ that can have a dimension $2$ bottom scalar component, and that contain an $SU(2)_R$ singlet with $R_2 = \pm 4$, are the ${\bf 6}_{-4}$ and ${\bar {\bf 6}}_{4}$. But we already saw that these contain an extra conserved supercurrent. Thus, pure $N=3$ theories cannot have any exactly marginal deformations.

The above analysis leaves open the possibility of having exactly marginal deformations that preserve a smaller amount of superconformal symmetry ($N=1$ or $N=2$). One cannot preserve $N=2$ by such deformations, since
exactly marginal deformations cannot modify the global symmetries of $N=2$ theories unless operators with spin $s \geq 2$ become conserved at the enhancement point, implying that the theory becomes free (see \cite{Beem:2014zpa}). It is possible that exactly marginal deformations preserving $N=1$ superconformal symmetry could exist, as is the case for the $N=4$ supersymmetric Yang-Mills theory. 

\subsection{$\mathbf{N=3}$ multiplets with global symmetries} \label{cvr}

If we have a global symmetry of some $N=3$ theory, it remains a global symmetry also when we view it as an $N=2$ theory. The bottom component of the global symmetry current multiplet of $N=2$ SCFTs is a scalar of dimension $2$, that is a triplet of $SU(2)_R$ and has $R_2=0$. By the same arguments as above, the bottom component of the $N=3$ multiplet must be a scalar with $\Delta_3=2$, and it
 should contain a triplet of $SU(2)_R$ with $R_2=0$ when supersymmetry  is reduced to $N=2$. Using again the unitarity constraints of appendix \ref{scaldim}, we find that the only possible R-charges of this bottom component are  ${\bf 6}_{-4}$, ${\bar {\bf 6}}_{4}$ and ${\bf 8}_0$. But we already saw that the first two cases contain an extra supercharge, and the latter is the $N=3$ energy-momentum multiplet that contains only the R-symmetry global currents. Thus, 
we conclude that pure $N=3$ theories cannot have global symmetries that are not R-symmetries.

\subsection{$\mathbf{N=3}$ multiplets with Coulomb branch operators}\label{cbos}

$N=3$ SCFTs may or may not have a moduli space. If they do, then since the only free representation in $N=3$ theories is a vector representation, the
low-energy theory at generic points on the moduli space must involve some number of these vector multiplets; in particular it should be called a Coulomb branch. Each $N=3$ vector multiplet contains six scalar fields, so a rank $r$ moduli space is labeled by $6r$ scalars.

In supersymmetric field theories (even with $N=1$ supersymmetry), moduli spaces may be labeled by the expectation values of some chiral operators (which are the bottom components of chiral multiplets). In this section
we examine the ``Coulomb branch operators'' (CBOs) in general $N=3$ SCFTs. These should be scalars which are lowest components of $N=3$ short superconformal multiplets. Since any $N=3$ theory is also an $N=2$ theory, we can use the information on CBOs in $N=2$ theories.

From the $N=2$ point of view, each $N=3$ vector multiplet decomposes into an $N=2$ vector multiplet and an $N=2$ hypermultiplet. Thus, the $N=3$ moduli space contains a $2r$-dimensional subspace which is an $N=2$ Coulomb branch (where only vector multiplets are turned on), and a $4r$-dimensional subspace which is an $N=2$ Higgs branch (where only hypermultiplets are turned on); all other points are mixed Coulomb-Higgs branches. In $N=2$ theories, a $2r$-dimensional Coulomb branch is labeled by $r$ $N=2$ chiral multiplets (often denoted by ${\cal E}_{{({R_2}/2)}(0,0)}$) whose bottom component is an $SU(2)_R$ singlet with 
$\Delta = |R_2| / 2$. These $r$ operators are believed to obey no relations in the $N=2$ chiral ring. The $N=3$ theory must therefore contain $r$ $N=3$ multiplets that contain such $N=2$ multiplets (including their lowest component). In
addition there must be Higgs-branch chiral operators from the point of view of the $N=2$ theory, that label the $4r$-dimensional Higgs branch; these are all have $R_2=0$. 

Moreover, we know that the $SU(3)_R$ symmetry rotates the $N=2$ vector multiplets into $N=2$ hypermultiplets
(this symmetry is generally broken on the moduli space, but we still know how it acts on the supercharges). Thus, every point on the Coulomb branch (on which the $SU(2)_R$ subgroup of $SU(3)_R$ is unbroken) can be rotated by an $SU(3)_R$ transformation into a point on the Higgs branch (where the $U(1)_{\tilde R}$ symmetry of the $N=2$ theory is unbroken).

An $N=3$ CBO containing an $N=2$ CBO could be either an $SU(3)_R$-singlet or a non-trivial representation. If it is a non-trivial representation, then the considerations of the previous paragraph imply that by an $SU(3)_R$ rotation the bottom component of the same $N=3$ multiplet should also contain $N=2$ operators labeling the Higgs branch, which have $R_2=0$. This limits the possible $N=3$ representations to have $SU(3)$ representations with weights $[a;0]$ (a symmetric product of $a$ ${\bf 3}$'s) and $R_3=-2a$, or their complex conjugates; these have dimension $\Delta = a$,
and obey the shortening condition \eqref{d2d4} on $N=3$ superconformal multiplets.
The case with $a=2$ is the one we discussed before that contains extra conserved supercharges, so such representations in pure $N=3$ theories must have $a \geq 3$. This class of operators appears also parameterizing the moduli space of $N=4$ theories (with larger $N=4$ multiplets that contain these $N=3$ multiplets); in that case the $a=2$ representation always appears, since it is part of the $N=4$ energy-momentum multiplet.

The bottom component could also be an $SU(3)_R$ singlet, as long as it has $|R_3| > 6$ and $\Delta = |R_3| / 6$. In this case we can use the fact that the $N=3$ superconformal algebra actually contains three separate $N=2$ superconformal algebras, each containing two of its three supercharges. The Coulomb branches from the point of view of the other $N=2$ subalgebras are part of the Higgs branch of the original $N=2$ subalgebra that we discussed, and should be labeled by appropriate $N=2$ chiral operators as well. But if we have an $N=3$ chiral multiplet that is an $SU(3)_R$ singlet then it is a chiral multiplet from the point of view of all $N=2$ subalgebras, contradicting the fact that it is supposed to be non-zero on the Coulomb branch but not on the Higgs branch of the moduli space. So we conclude that such multiplets (containing $N=2$ CBOs) should not exist.

So the moduli space of $N=3$ theories must be (at least partly) labeled by $r$ representations of the type mentioned above, with some integers $a=3,4,\cdots$. The fact that from the $N=2$ point of view there are no relations between the $r$ CBOs implies that there are no constraints (in the sense of constraints on the expectation values of the corresponding operators in supersymmetric vacua) relating the symmetric products of the corresponding $N=3$ multiplets, though there would always be other constraints relating the multiplets (to ensure that the moduli space is just $6r$-dimensional). Of course there would generally be an infinite number of $N=3$ CBOs, but only $r$ of them should be independent in the sense mentioned here. 

Apriori there could be extra $N=3$ CBOs that do not contain $N=2$ CBOs, and that label regions of the moduli space that do not contain the $N=2$ Coulomb branch. Our arguments above imply that there are no constraints relating the $N=2$ CBOs inside the $N=3$ CBOs, and also the operators in the same $SU(3)_R$ representations that have $R_2=0$ and are highest and lowest weights of $SU(2)_R$ (these are related by permutations of the three supercharges to the $N=2$ CBOs). We conjecture that these $N=3$ CBOs that contain $N=2$ CBOs are enough to parameterize the full moduli space. However, we could not prove this, since there could be constraints relating the three different types of operators in the $N=3$ CBOs (that correspond to the Coulomb branches of the different $N=2$ subgroups).

There is a conjecture \cite{Tachikawa:2013kta} that the chiral ring of $N=2$ theories is freely-generated; this is true in all known $N=2$ theories. If it is true, then any $N=3$ SCFT that contains a multiplet containing an $N=2$ chiral (Coulomb-branch) multiplet must have a non-trivial moduli space, parameterized as above. However, even if this conjecture is true, we cannot conclude that all $N=3$ theories have a moduli space, since there is no obvious reason why every $N=3$ theory should contain such a multiplet (unlike $N=4$ SCFTs which, as discussed above, always contain such a multiplet).

It would be interesting to find further constraints on the Coulomb branches and the dimensions of CBOs of pure $N=3$ theories. The leading irrelevant operators on the Coulomb branch of $N=3$ theories were studied using an $N=3$-preserving superspace \cite{Galperin:1985uw,Galperin:1984bu} in \cite{Buchbinder:2004rj,Buchbinder:2011zu}, and it was found that they are the same as the ones appearing on the Coulomb branch of the $N=4$ supersymmetric Yang-Mills theory, though their coefficient may be different. Higher irrelevant operators on the Coulomb branch of $N=3$ theories are presumably more general than the ones appearing in the $N=4$ theory, and it would be interesting to analyze their properties.

\bigskip
\noindent{\bf Acknowledgments}

We would like to thank Marco Baggio, Nikolay Bobev, George Papadopoulos and Igor Samsonov for useful discussions, and Zohar Komargodski for many useful discussions and collaboration on parts of this project.
This work was supported in part by an Israel Science Foundation center for excellence grant, by the I-CORE program of the Planning and Budgeting Committee and the Israel Science Foundation (grant number 1937/12), by the Minerva foundation with funding from the Federal German Ministry for Education and Research, by a Henri Gutwirth award from the Henri Gutwirth Fund for the Promotion of Research, and by the ISF within the ISF-UGC joint research program framework (grant no. 1200/14). OA is the Samuel Sebba Professorial Chair of Pure and Applied Physics.

\appendix

\section{No relevant deformations of $\mathbf{N=3}$ SCFTs}\label{relevant}

One possible way to obtain an $N=3$ theory would be to deform an $N=4$ theory by a relevant deformation that breaks $N=4$ but leaves $N=3$ intact; this would lead to an $N=3$ theory in the infrared limit
 (our discussion in section \ref{emd} shows that one cannot break $N=4$ to $N=3$ with an exactly marginal deformation).
In this appendix we show that this is impossible by proving a more general statement: any $N=3$ theory (including $N=4$ theories)
cannot have a relevant deformation preserving the full $N=3$ supersymmetry.

To have such a deformation there
 should be an $N = 3$ multiplet with a top component that is a scalar and that has scaling dimension
$\Delta < 4$. Relevant deformations of $N=2$ SCFTs that preserve supersymmetry can either lie
in conserved current multiplets or in (conjugate) chiral multiplets. We have already shown in section \ref{cvr} that pure
$N=3$ theories cannot contain conserved current multiplets, and that in the $N=4$ SYM theory the only conserved currents are in the energy-momentum multiplet that does not contain $N=3$-preserving relevant deformations, so an $N=3$ multiplet containing a relevant
deformation must contain an $N=2$ (conjugate) chiral multiplet.

Let us restrict to the case of the $N=2$ chiral multiplet
without losing generality.
In the $N=2$ chiral multiplet, the top component arises by the action of four $Q$ supercharges on the
bottom component. In the $N=3$ multiplet, the top component must then arise by the action
of at least four supercharges on the bottom component. But since it is a scalar with $\Delta < 4$,
unitarity (the non-existence of fermionic operators with $\Delta < \frac{3}{2}$ and of bosonic
operators with $\Delta < 1$) implies that the top component must arise from the action of exactly four supercharges
(which are all $Q$'s and not ${\bar Q}$'s).
We then have two different cases:
\begin{itemize}
\item{1.} The bottom component is a singlet of $SU(3)_R$. In this case after acting with
four supercharges we necessarily obtain an operator charged under $SU(3)_R$ (since it
has triality one). But this means that when the supersymmetry is decomposed from $N = 3$ to $N = 2$, 
we would get an
$N = 2$ multiplet that has a scalar top component with  $\Delta < 4$ and that is charged under $SU(2)_R$.
But there is no such $N=2$ superconformal multiplet.
\item{2.} The bottom component is charged under $SU(3)_R$. As we want the scaling dimension of
the bottom component to be $\Delta < 2$, we are severely restricted by the unitarity restrictions
of appendix \ref{scaldim}. The only $SU(3)_R$-charged operators that satisfy them with $\Delta < 2$
are scalars in the ${\bf 3}_{-2}$ or ${\bar {\bf 3}}_2$ representations, with $\Delta=1$. 
However, this corresponds to a free $N = 3$ vector multiplet, that is not present in an
interacting theory.
\end{itemize}

Thus, we conclude that $N=3$ theories cannot have SUSY-preserving relevant deformations, and, in
particular, that one cannot obtain an $N=3$ theory by a relevant deformation of an $N=4$ theory.
Note that $N=3$ SCFTs do always have a relevant deformation that preserves $N=2$ supersymmetry,
since their energy-momentum multiplet contains an $N=2$ global current multiplet, which contains
such a deformation.

\section{Unitarity restrictions on the bottom component of a multiplet} \label{scaldim}

The scaling dimension $\Delta_0$ of the bottom component of an $N=3$ superconformal multiplet is constrained according to equations (6a)-(6d) of \cite{Dobrev} (see also \cite{Minwalla}). Taking the bottom component to have spin $(j_1,j_2)$, $U(1)_R$ charge $R_3$, and to sit in the $SU(3)_R$ representation $[a;b]$ (labeling the representation by its weights), unitarity constrains generic multiplets by
\begin{align}
& \Delta_0 \geq d_1 = 2 + 2j_1 + \frac{4a+2b}{3} + \frac{R_3}{6},\label{d13}\\ 
& \Delta_0 \geq d_3 = 2 + 2j_2 + \frac{2a+4b}{3} - \frac{R_3}{6}	.\label{d24} 
\end{align}
In addition there are a few special cases:

{\bf 1.} If the bottom component is a scalar, then in addition we have a unitary representation when one of the following three conditions is satisfied:
\begin{align}
& \Delta_0 = d_4 = \frac{2a+4b}{3} - \frac{R_3}{6} \quad \mathrm{if} \quad d_4 \geq d_1, \label{d4}\\
& \Delta_0 =  d_2 = \frac{4a+2b}{3}+\frac{R_3}{6} \quad \mathrm{if} \quad d_2 \geq d_3, \label{d2}\\
& \Delta_0 = d_2 \quad \mathrm{if} \quad d_2 = d_4. \label{d2d4}
\end{align}
An important example of a multiplet that satisfies \eqref{d2d4} is the free vector multiplet $[1; 0]_{-2}={\bf 3}_{-2}$.

{\bf 2.} If the bottom component has spin $(j,0)$, then the unitarity conditions are also satisfied when
\begin{equation}
\Delta_0 = d_4 = -2j + \frac{2a+4b}{3} - \frac{R_3}{6} \quad \mathrm{if} \quad d_4 \geq d_1.
\end{equation}

{\bf 3.} If the bottom component has spin $(0,j)$, then the unitarity conditions are also satisfied when
\begin{equation}
\Delta_0 =  d_2 = -2j + \frac{4a+2b}{3}+\frac{R_3}{6} \quad \mathrm{if} \quad d_2 \geq d_3. \label{dj2}
\end{equation}

Whenever one of \eqref{d4}--\eqref{dj2} is satisfied, or we have an equality in \eqref{d13} or \eqref{d24}, we have a short multiplet with less states than a generic long multiplet. All the multiplets we discuss in this paper will be of this type.

\end{document}